\documentclass[aps,pre,preprint,superscriptaddress,amsmath,amssymb,nofootinbib]{revtex4-2}

\usepackage{amsmath}
\usepackage{graphicx}
\usepackage{amssymb}
\usepackage{dcolumn}
\usepackage{bm}
\usepackage{color}
\usepackage{float}
\usepackage{upgreek}
\usepackage{gensymb}
\usepackage[final]{hyperref} 
\hypersetup{
	colorlinks=true,       
	linkcolor=blue,        
	citecolor=blue,        
	filecolor=magenta,     
	urlcolor=blue
}
\usepackage{upgreek}
\usepackage[normalem]{ulem}

\makeatletter

\begin{document}
\title{Mass-based separation of active Brownian particles in an asymmetric channel}
	
\author{Narender Khatri}\thanks{Electronic mail: K.Narender@lmu.de}
\affiliation{Department of Veterinary Sciences, Institute for Infectious Diseases and Zoonoses, Ludwig-Maximilians-Universitaet Munich, 80539, Germany}


\date{\today}
	
\begin{abstract}
Inertial effects should be considered for micro- and nano-swimmers moving in a low-density medium confined by irregular structures that create entropic barriers, where viscous effects are no longer paramount. Here, we present a separation mechanism of self-propelled particles in a two-dimensional asymmetric channel, which leads to the drift of particles of different masses in opposite directions. In particular, this mechanism is based on the combined action of the spatial asymmetry of the channel structure, the temporal asymmetry inherent in particles dynamics, and an external static force. This work is relevant for potential applications that can be found in the development of lab-on-a-chip devices and artificial channels for separating particles of different masses.  
\end{abstract}
	
\maketitle

\section{Introduction}

For the most part, active matter systems consist of microscopic or sub-microscopic active agents, biological as well as synthetic, that take the free energy from their environments and convert it to a persistent motion under nonequilibrium conditions \cite{LP2009,Ramaswamy@ARCMP:2010,Cates@RPP:2012,2013perspective,A13, Palacci_et_al@Science:2013, Sengupta_et_al@JACS:2013, Wang2015,Elgeti_et_al@RPP:2015,Zoettl2016,Bechinger_et_al@RMP:2016,Illien2017,Archer_et_al@AS:2018,GK19,Gaspard_Kapral@Research:2020,Gompper_et_al@JPCM:2020,khatri2023clustering}. Nonequilibrium dynamics of such particles, also termed as micro- and nano-swimmers, are often described by the overdamped Langevin or continuum models, where the viscous force is paramount and inertial effects can be neglected. However, in low-density media, inertial effects become relevant due to reduced viscous force, and a body of research \cite{Lowen-per2020} describes the new phenomena as a result of their inclusion. Some examples where inertial effects are important include active particles motion in low-density media such as gases \cite{Sprenger_et_al@PRE:2021,Caprini_et_al@PCCP:2022,Leoni_et_al@PRR:2020}, plasmas \cite{Morfill_Ivlev@RMP:2009,Arkar_et_al@Molecules:2021,Li_et_al@PRR:2023} and superfluids \cite{Kolmakov_Aranson@PRR:2021}, vibrobots \cite{Scholz2016,Torres2016,Scholz_et_al@NC:2018}, active aerosols \cite{Rohde_et_al@OE:2022}, and systems that support a temperature gradient across coexisting phases \cite{mandal2019}. Inertial active particles are called micro- and nano-flyers rather than swimmers \cite{Lowen-per2020,Khatri_Kapral@JCP:2023}.  

The active matter very often exists in the form of a mixture with a wide distribution of sizes, masses, motilities, chiralities, and shapes. Separating wanted active particles from unwanted ones is of great fundamental and nanotechnological importance for various branches of science and engineering. So far, much effort has been devoted to the motility-based \cite{Maggi_et_al@SM:2013,Berdakin_et_al@PRE:2013,Costanzo_et_al@EL:2014,Zaferani_et_al@PNAS:2018} and chirality-based \cite{Barois_et_al@PRL:2020,Lin_et_al@JCP:2020,Doi_Makino@PF:2016,Ai_et_al@SM:2018,Scholz_et_al2@NC:2018,Reichhardt_Reichhardt@PRE:2013,Mijalkov_Volpe@SM:2013,Chen_Ai@JCP:2015,Ai_et_al@SM:2015} separation of particles. However, the mass-based separation of particles is of utmost importance for micro- and nano-robotics, environmental monitoring, nanotechnological applications, etc. In particular, separating particles based on their masses can be very challenging because, typically, identically-sized particles can have different masses. 

In this paper, we present a mass-based separation mechanism of active particles in a two-dimensional asymmetric channel. The irregular shape of the considered channel structure gives rise to entropic barriers \cite{Zwanzig@JCP:1992,Reguera_Rubi@PRE:2001,Kalinay_Percus@PRE:2006,Reguera_et_al@PRL:2012,Motz_et_al@JCP:2014,Khatri_Burada@PRE:2020,Khatri_Burada@PRE:2021,Gupta_Burada@PRE:2023,Galajda_et_al@JB:2007} that significantly influence the diffusive behavior of particles. As well, the spatial asymmetry of the channel can induce active directed transport of particles in the absence of external forces, i.e., so-called entropic rectification \cite{Ghosh_et_al@PRL:2013,Reichhardt_Reichhardt@ARCMP:2017,Khatri_Kapral@JCP:2023}. 
It should be noted that we have reported how inertia affects the entropic rectification and diffusion of particles using the same channel in Ref.~\cite{Khatri_Kapral@JCP:2023}. 
The results that follow show how particles of different masses can be separated in opposite directions in the presence of an external static force by purely entropic means. 

The rest of the paper is organized as follows: in Sec.~\ref{sec: Model}, we introduce the minimal underdamped Langevin model used to describe the dynamics of active particles in a two-dimensional asymmetric channel. Section~\ref{Sec: Results} discusses the results of the mass-based separation of particles, and the main conclusions are given in Sec.~\ref{Sec: Conclusions}.

\section{Model}
\label{sec: Model}

\begin{figure}[hbt]
\centering
\resizebox{0.7\columnwidth}{!}{%
\includegraphics[scale = 1.0]{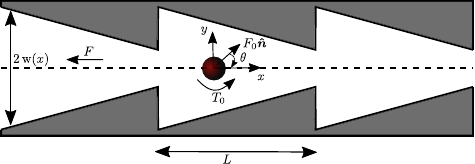}}
\caption{Schematic illustration of a two-dimensional (triangular-shaped) asymmetric channel, described by Eq.~(\ref{eq:wall}), with periodicity $L$ confining an active Brownian particle of mass $M$ and moment of inertia $I$, which is subjected to an external homogeneous constant force $\boldsymbol{F} = - F \boldsymbol{\hat{x}}$ along the negative $x$-direction. The local width of the channel $2 \, {\rm w} (x)$, active force $\boldsymbol{F_0} = F_0 \boldsymbol{\hat{n}}$, angle $\theta$, and active torque $T_0$ are indicated.}
\label{fig:Model}
\end{figure}
We consider a dilute suspension of active particles in a dissipative medium constrained to move in a two-dimensional asymmetric channel with periodicity $L$ (see Fig.~\ref{fig:Model}). We neglect the direct interactions among particles under the assumption of sufficiently small particle density. The dynamics of an active particle with position $\boldsymbol{r}  = (x, y)$, orientation $\boldsymbol{\hat{n}} = (\cos{ \theta }, \sin{\theta })$, mass $M$, and moment of inertia $I$, driven by an external homogeneous static force $\boldsymbol{F} = - F \boldsymbol{\hat{x}}$ acting along the negative $x$-direction, is described by the coupled Langevin equations as  
\begin{equation}\label{Eq:Langevin1}
 \begin{aligned}
    M \ddot{\bm{r}}(t) &= -\gamma_t \Dot{\bm{r}}(t) - F \boldsymbol{\hat{x}}  + F_0 \boldsymbol{\hat{n}} (t) + \gamma_t \sqrt{2 D_t}~ \boldsymbol{\xi} (t), \\
    I \Ddot{\theta}(t) &= -\gamma_r \Dot{\theta}(t) + T_0 + \gamma_r \sqrt{2 D_r} ~ \zeta (t), 
    \end{aligned}
\end{equation} 
where $\gamma_t$ and $\gamma_r$ denote the translational and rotational friction coefficients, $F_0$ and $T_0$ are the active force and torque originating from the active processes that lead to propulsion \cite{Bechinger_et_al@RMP:2016, Gaspard_Kapral@Research:2020,Illien2017}, and $D_t$ and $D_r$ are the translational and rotational diffusion constants, respectively.
 Note that $F_0$ and $T_0$ remain constant with time, and the orientational correlation function of the active force is given by $\langle \boldsymbol{\hat{n}} (t).\boldsymbol{\hat{n}} (0) \rangle = \cos(\omega t) e^{-D_r (t \, - \, \tau_\omega(1 \, - \, e^{-t /\tau_\omega}))}$, where $\omega = T_{0}/\gamma_r$ and $\tau_\omega = I/\gamma_r$ denote the angular velocity and the angular velocity relaxation time, respectively \cite{Lowen-per2020}.
 The translational and rotational Brownian fluctuations from the surrounding medium are modeled by the Gaussian white noise terms $\boldsymbol{\xi} (t)$ and $\zeta (t)$, respectively, with zero mean and unit variances given by $\langle \boldsymbol{\xi} (t) \otimes \boldsymbol{\xi}^{\mathrm{T}} (t')\rangle = \delta (t - t') \mathrm{\bf 1}$ and $\langle \zeta (t) \zeta (t') \rangle = \delta (t - t')$, where $\mathrm{\bf 1}$ is the identity matrix. 
As is well known, the set of coupled Langevin equations~(\ref{Eq:Langevin1}) serves as models for diverse active systems, including systems where the noise is athermal. For systems subject to athermal noise \cite{Berg@Book:2004,Fily_Marchetti@PRL:2012,Palacci_et_al@Science:2013,Marchetti_et_al@COCIS:2016,Khatri_Kapral@JCP:2023,Scholz_et_al@NC:2018}, $\{\gamma_t, \gamma_r, D_t, D_r \}$ are treated as independent parameters, and $D_t$ and $D_r$ control the strength of the translational and rotational noises, respectively. However, the diffusion and friction constants are related by Stokes-Einstein relations, $D_t = k_B T/ \gamma_t$ and $D_r = k_B T/ \gamma_r$, for systems with thermal noises that satisfy the fluctuation-dissipation relation. 
In this setup, $\boldsymbol{F}$ can be imposed by applying electric, magnetic, and acoustic fields, or a combination of them, if particles are sensitive to these fields \cite{Toschi_Sega@Book:2019}. For instance, Loget and Kuhn \cite{Loget_Kuhn@NC:2011} have added an external force to particles constructed from conducting materials by applying an electric field. Since $\boldsymbol{F}$ is considered to be homogeneous, it cannot cause the hydrodynamic flow.
The generalization of the considered set of coupled Langevin equations~(\ref{Eq:Langevin1}) to a more general set of coupled Langevin equations that accounts for asymmetric particles has been given in Ref.~\cite{Martins_Wittkowski@PRE:2022}. Using fluctuating chemohydrodynamics \cite{Gaspard_Kapral@JCP:2018} and molecular theory \cite{Robertson_et_al@JCP:2020}, a set of general underdamped coupled Langevin equations has been derived for chemically-powered colloids, where active force and torque expressions are given. 

For the two-dimensional asymmetric and spatially periodic channel with periodicity $L$ depicted in Fig.~\ref{fig:Model}, the channel walls at position $x$ are described by 
\begin{equation}\label{eq:wall}
{\rm w}_u (x) = \begin{cases}
{\rm w}_\mathrm{min}, & x = 0,\\
{\rm w}_\mathrm{max} - ({\rm w}_\mathrm{max} - {\rm w}_\mathrm{min}) \frac{x}{L}, & 0 < x\leq L,
\end{cases}
\end{equation} 
where ${\rm w}_u (x)$ and ${\rm w}_l (x) = - {\rm w}_u (x)$ are the upper and lower channel walls, ${\rm w}_\mathrm{min}$ and ${\rm w}_\mathrm{max}$ are the minimum and maximum half-widths of the channel, respectively, and $2 \, {\rm w}(x) = {\rm w}_u (x) - {\rm w}_l (x)$ is the local width of the channel. The dimensionless parameter $\epsilon = {\rm w}_\mathrm{min}/{\rm w}_\mathrm{max}$ defines the aspect ratio of the channel, where we choose ${\rm w}_\mathrm{max} = L$  throughout the work. Note that $\epsilon = 1$ corresponds to a flat channel. 

The collisional dynamics of the particle at the channel walls is modeled by sliding-reflecting boundary conditions \cite{Khatri_Kapral@JCP:2023,Ghosh_et_al@PRL:2013,Khatri_Burada@PRE:2022,Reichhardt_Reichhardt@ARCMP:2017} as follows: the translational velocity $\Dot{\bm{r}}$ of the particle is elastically reflected \cite{Khatri_Burada@JCP:2019,Khatri_Burada@JSM:2021}, and its orientation $\boldsymbol{\hat{n}}$ is unchanged during the collision. Consequently, the active force $\boldsymbol{F_0} = F_0 \boldsymbol{\hat{n}}$ keeps pointing in the same direction; hence, the particle slides along the channel wall until a fluctuation in the orientational vector $\boldsymbol{\hat{n}}$ redirects it towards the interior of the channel. While our consideration is restricted only to this collision mechanism, it should be noted that the particle's orientation will change during the collision if it interacts with the channel wall through rough sphere collisions as well as its interaction with the solvent particles. The latter collision mechanism depends on the scale of channel wall roughness. 

Specifically, when the dynamics takes place in a periodic channel, the motion of an active particle governed by Eq.~(\ref{Eq:Langevin1}) is often determined by the linear velocity relaxation time $\tau_v = M/\gamma_t$, angular velocity relaxation time $\tau_\omega = I/\gamma_r$, reorientation time $\tau_r = 1/D_r$, spinning orientation time $\tau_s = \gamma_r/|T_0|$, and characteristic diffusion time $\tau = L^2/D_t$ \cite{note:Bulk}. It is convenient to use a dimensionless description \cite{Khatri_Kapral@JCP:2023} where length variables are scaled by $L$, $\bm{r}' = \boldsymbol{r}/L$, and time by $\tau$, $t' = t/\tau$. In the following, we shall dispense with prime symbols for better readability. The Eq.~(\ref{Eq:Langevin1}) in dimensionless form reads 
\begin{equation}\label{Eq:Langevin2}
 \begin{aligned}
    M^* \Ddot{\bm{r}}(t) &= -{\Dot{\bm{r}}}(t) - f \bm{\hat{x}} + f_0 \boldsymbol{\hat{n}} (t) + \sqrt{2} ~\boldsymbol{\xi} (t), \\
    I^* \Ddot{\theta}(t) &= -\Dot{\theta}(t) + t_0 +  \sqrt{2 \alpha} ~ \zeta (t). 
    \end{aligned}
\end{equation}  
Here the dimensionless parameters are given by $M^* = \tau_v/\tau = M D_t /(\gamma_t L^2)$, $I^* = \tau_\omega/\tau = I D_t/(\gamma_r L^2)$, $f = F L/(D_t \gamma_t)$, $f_0 = F_0 L/(D_t \gamma_t)$, $t_0 =  T_0 L^2/(D_t \gamma_r)$, and $\alpha = D_r \tau$. Notably, one could have used other choices of dimensionless units to find the number of independent parameters in the system \cite{Lowen-per2020}.  

In the following, we present results for the mass-based separation of particles in opposite directions. The observable of foremost interest for separating particles is the stationary average velocity along the principal axis of the channel ($x$-direction). Unfortunately, the explicit analytical expression for the average velocity cannot be obtained for the considered system due to the extreme difficulty of analytically solving the Fokker-Planck equations \cite{Ghosh_et_al@PRL:2013,Reichhardt_Reichhardt@ARCMP:2017}. For this reason, the average velocity is calculated from simulations of the coupled Langevin equations~(\ref{Eq:Langevin2}) in the channel~\cite{note:sim}. As an initial condition at time $t = 0$, the mixture of particles, with random orientations, of various masses was placed in a single cell of the channel located between $x = 0$ and $x = 1$, and the results are obtained by averaging over an ensemble of $10^{4}$ stochastic trajectories. 

\section{Particle separation}\label{Sec: Results}

\begin{figure}[htb!]
\centering
\resizebox{0.7\columnwidth}{!}{%
\includegraphics[scale = 1.0]{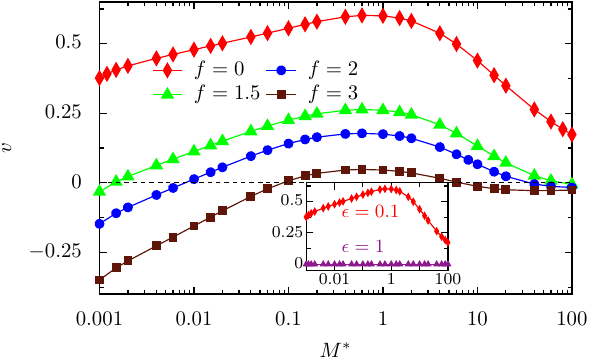}}
\caption{The plot of the average velocity $v$ versus $M^*$ for different values of the static force $f$. Here and below, the statistical error for $v$ is smaller than the symbol sizes, and the solid lines are guides to the eye. The set parameters are: $I^* = 0.001, \alpha =  0.1, f_0 = 5, \epsilon = 0.1, \mathrm{and}~t_0 = 0$. For $f = 0$, the inset plots $v$ versus $M^*$ for $\epsilon = 0.1$ and $\epsilon = 1$.}
\label{fig:Graph2}
\end{figure}
\begin{figure}[htb!]
\centering
\resizebox{0.7\columnwidth}{!}{%
\includegraphics[scale = 1.0]{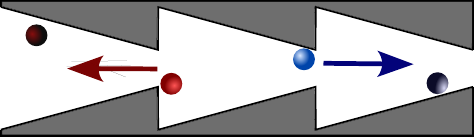}}
\caption{Schematic illustration of separating particles of the same size but different masses, placed initially in the center, in opposite directions. Under the combined action of the static force, shape of the channel structure, and active motion of particles, lighter particles are following the static force, i.e., are moving towards the left side of the channel, whereas heavier particles are drifting towards the right.}
\label{fig:Graph3}
\end{figure}

One of the best ways to separate particles is to move them in opposite directions, which in our setup is achieved by applying a small static force $f$ that controls the transport direction. The average velocity $v$ is plotted as a function of $M^*$ in Fig.~\ref{fig:Graph2} for several values of the static force $f$. In the absence of $f$, all particles move towards the right side of the channel ($v > 0$) due to their rectification ascribed by the spatial asymmetry of the channel structure and the temporal asymmetry inherent in the particle dynamics. The rectification strongly depends on $M^*$, and $v$ has a peak at an optimal mass $M^*_{\rm op}$. We have reported the dependence of $M^*_{\rm op}$ on the rotational diffusion rate $\alpha$, active force $f_0$, aspect ratio of the channel $\epsilon$, and active torque $t_0$ in Ref.~\cite{Khatri_Kapral@JCP:2023}. In the presence of small $f$, particles lighter than a given threshold follow $f$, i.e., move towards the left ($v < 0$), whereas heavier particles than this drift towards the right. However, only particles having mass within a band of $M^*$ move towards the right on slightly increasing $f$, and $v$ exhibits a valley behavior as a function of $M^*$ for higher $M^*$ values because, in this regime, effects due to the static force dominate over rectification effects. It is worth mentioning that $M^*_{\rm op}$ is found to be independent of the static force and moment of inertia. As expected, when $M^* \to \infty$, i.e., in the strongly underdamped limit, inertia dominates over the self-propulsion and static force, resulting in $v$ tending to zero \cite{note:LargeM}. The mass-based separation of particles of the same size is illustrated schematically in Fig.~\ref{fig:Graph3}.

When compared to $\epsilon = 0.1$, in the absence of the static force $f$, the average velocity $v$ of particles of any mass $M^*$ is zero for $\epsilon = 1$ due to the spatial symmetry of the channel structure (see the inset of Fig.~\ref{fig:Graph2}). Since, in the latter case, the rectification of particles is zero ($v = 0$ for $f = 0$), these particles cannot move in the opposite direction of $f$. Therefore, in our setup, the flat channel cannot lead to the mass-based separation of particles.

\begin{figure}[htb!]
\centering
\resizebox{0.7\columnwidth}{!}{%
\includegraphics[scale = 1.0]{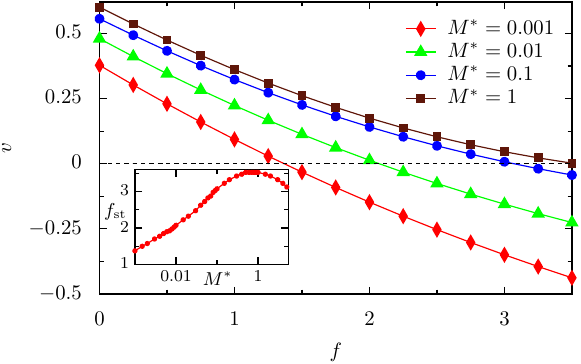}}
\caption{The plot of the average velocity $v$ versus the static force $f$ for different values of $M^*$. The chosen parameters are the same as in Fig.~\ref{fig:Graph2}. The inset plots the dependence of the stall force $f_{\mathrm {st}}$ on $M^*$.}
\label{fig:Graph4}
\end{figure}
The dependence of the average velocity $v$ on the static force $f$ is depicted in Fig.~\ref{fig:Graph4} for different values of $M^*$. As is reflected in this figure, the mass-based separation of particles can be effectively controlled by suitably tuning $f$. For instance, by choosing $f = 2.5$, lighter particles of mass $M^* = 0.01$ move towards the left with $v \approx -0.1$, whereas heavier particles of mass $M^* = 0.1$ drift towards the right with $v \approx 0.1$. In particular, there exists a stall force $f_{\mathrm {st}}$ for which $v$ is zero. For $f  < f_{\mathrm {st}}$, particles of mass $M^*$ move towards the right; whereas the same particles drift towards the left for $f  > f_{\mathrm {st}}$. $f_{\mathrm {st}}$ shows a nonmonotonic dependence on $M^*$ with the appearance of a peak at the optimal mass $M^*_{\rm op}$ (see the inset). More interestingly, the nonmonotonic dependence of $f_{\mathrm {st}}$ on $M^*$ provides that we obtain a smart device with a fixed shape that can be used to continuously separate particles of any mass by suitably changing merely $f$.

\begin{figure}[htb!]
\centering
\resizebox{0.5\columnwidth}{!}{%
\includegraphics[scale = 1.0]{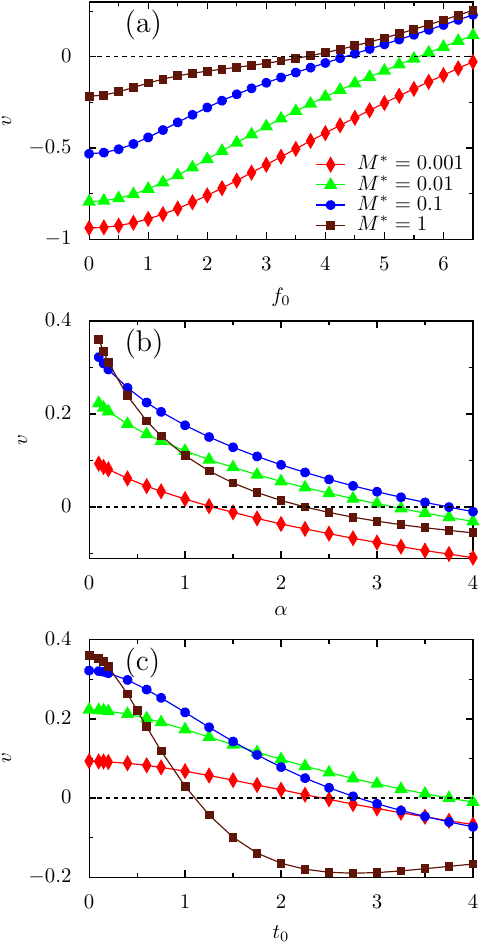}}
\caption{The plot of the average velocity $v$ dependence on the self-propulsion force $f_0$ (a), rotational diffusion rate $\alpha$ (b), and active torque $t_0$ (c) for different values of $M^*$. For panel (a), $\alpha = 0.1, t_0 = 0, I^* = 0.001, \epsilon = 0.1, \mathrm{and}~f = 2.5$, for panel (b), $f_0 = 5, t_0 = 0, I^* = 0.001, \epsilon = 0.1, \mathrm{and}~f = 1$, and for panel (c), $f_0 = 5, \alpha = 0.1, I^* = 0.001, \epsilon = 0.1, \mathrm{and}~f = 1$.}
\label{fig:Graph5}
\end{figure}
The dependence of the average velocity $v$ on the self-propulsion force $f_0$, rotational diffusion rate $\alpha$, and active torque $t_0$ is shown in Fig.~\ref{fig:Graph5} for different values of $M^*$. For fixed values of $\alpha, t_0, \mathrm{and}~f$, the mass-based separation of particles can be controlled by suitably tuning $f_0$ (see panel (a)). As is reflected from panels (b) and (c), the separation of particles can also be tuned by values of $\alpha$ and $t_0$. We have verified that $v$ is independent of the sign of $t_0$ because our channel is symmetric about its principal axis, i.e., it has top-down symmetry. In particular, as $f_0 \to 0$, $\alpha \to \infty$, or $t_0 \to \infty$, the motion of self-propelled particles approaches passive Brownian motion; thus, the rectification effect tends to vanish, and as a consequence, all particles follow the static force.  

\section{Concluding remarks}\label{Sec: Conclusions}

To summarize, the paper presented a mechanism to separate self-propelled particles of different masses by purely entropic means. This mechanism relies on the entropic rectification of particles caused by the spatial asymmetry of the channel structure and the temporal asymmetry inherent in their dynamics. In particular, the entropic rectification strongly depends on the mass of particles. An external static force can overcome the entropic rectification effect and lead to the mass-based separation of particles in opposite directions. Furthermore, the separation of particles can also be tuned by the self-propulsion force, rotational diffusion rate, and active torque. Viewed in the context, the presented mechanism has the potential to be experimentally implemented in asymmetric narrow channels and micropores, where entropic effects are prominent, for the separation of particles based on their masses. In the future, applying the mean-field methods to obtain the mass-based separation of particles in asymmetric channels will be insightful.\\



\section{Acknowledgment}

I would like to thank Prof. Dr. Raymond Kapral for some valuable discussions. This work was supported in part by the Natural Sciences and Engineering Research Council (NSERC) of Canada and Compute Canada (\href{https://www.computecanada.ca/}{www.computecanada.ca}).

\section{Conflict of Interest}
The author has no conflicts to disclose.

\section{Data availability}

The data that support the findings of this study are available from the author upon reasonable request.

\appendix
\section{Choice of parameters}\label{Sec: Appendix}

The presented mass-based separation mechanism can be applied to a wide variety of physical systems whose active agents are self-propelled by various mechanisms and are subject to either thermal or athermal noise. The asymmetric channel that we considered can be prepared by microprinting on a substrate. It is instructive to provide an estimate of parameters in real units, which is very useful for experimentalists. One should note that the choice of parameters places limits on physical systems. 

A class of systems that may be of interest are active aerosols~\cite{Rohde_et_al@OE:2022,Bakanov1987,Chernyak2001}, where our setup can lead to the mass-based separation of such particles in opposite directions. As an example, consider a mixture of identically-sized active particles of radii $a \sim 200~ \mathrm{nm}$ but different masses lie in the range $M \sim 10^{-16}-10^{-15} ~\mathrm{kg}$ with moment of inertia $I \sim  10^{-31} ~\mathrm{kg~m^2}$ in air at room temperature and pressure $p=10^4-10^5~\rm{Pa}$. The translational and rotational friction coefficients are given by $\gamma_t \sim 10^{-11} ~\mathrm{kg/s}$ and $\gamma_r \sim 10^{-24} ~\mathrm{kg~m^2/s}$, respectively, corresponding to the viscosity of the medium $\eta \sim 10^{-5}~\mathrm{kg/(m~s)}$ (independent of pressure). The active and external static forces lie in the range $F_0 \sim 0.1- 1 ~\mathrm{pN}$ and $F \sim 0.1- 1 ~\mathrm{pN}$, respectively, and the active torque is taken to be zero. The asymmetric channel parameters are fixed as ${\rm w}_\mathrm{max} = L = 10~\mu{\rm m}$ and $\epsilon =0.1$.

The Stokes-Einstein relations, hold for thermal noise, provide $D_t=k_BT/\gamma_t \sim 10^{-9}~\mathrm{m^2/s}$ and $D_r=k_BT/\gamma_r \sim 10^{3}~\mathrm{s^{-1}}$. Correspondingly, in dimensionless units, we obtain $\tau \sim 0.1 ~{\rm s}$, $M^* \sim 10^{-4}-10^{-3}$, $I^* \sim 10^{-6}$, $f_0 \sim 10^{2}-10^{3}$, $f \sim 10^{2}-10^{3}$, and $\alpha \sim 10^2$. From these parameters, we can see that the separation of particles cannot be achieved because the regime where inertial effects play a significant role cannot be accessed. 

For athermal noise, select $D_t \sim 10^{-8}-10^{-6}~\mathrm{m^2/s}$ and $D_r\sim 10^{3}~\mathrm{s^{-1}}$. Then, we obtain $\tau \sim 10^{-4}-10^{-2} ~{\rm s}$, $M^* \sim 0.001-1$, $I^* \sim 10^{-5}-10^{-3}$, $f_0 \sim 0.1-100$, $f \sim 0.1-100$, and $\alpha \sim 0.1-10$. The persistence length and P\'eclet number lie in the range $l_{p} = v_0/D_r \sim 10-100~\mu{\rm m}$ and $Pe = v_0/ \sqrt{D_t D_r} \sim 0.3-30$, respectively. Under these conditions, the separation of particles can be achieved in opposite directions.       



\bibliography{APS} 

\begin{thebibliography}{72}
\bibitem[{not({\natexlab{a}})}]{note:Bulk}
\bibinfo{note}{In a bulk medium, inertial effects dominate frictional effects
  for times $t \ll \tau_v$ for translational motion, while for orientational
  motion, they are significant for times $t \ll \tau_\omega$.}
\bibitem[{not({\natexlab{c}})}]{note:LargeM}
\bibinfo{note}{To better illustrate the trends of $v$, we have included results
  for very large $M^{*}$, i.e., the strongly underdamped limit; however, many
  physical systems may not access this limit.}
\end{thebibliography}


	
\end{document}